\documentclass[twoside,slac_one]{revtex4}
\usepackage{graphicx}
\usepackage{fancyhdr}
\usepackage{amsmath} % American Mathematics Society standards
\usepackage{bm}% bold math
\usepackage{amsxtra}
\usepackage{amssymb}
\usepackage{amsthm}
\usepackage{latexsym}
\usepackage{lscape}
\usepackage{atlasphysics}
\usepackage{lldefs}

\newcommand{\LimitElectronExpected}{1.70}
\newcommand{\LimitElectron}{1.70}

\newcommand{\LimitMuonExpected}{1.61}
\newcommand{\LimitMuon}{1.61}

\newcommand{\LimitCombinedExpected}{1.83}
\newcommand{\LimitCombined}{1.83}

\newcommand{\LimitElectronExpectedG}{1.50}
\newcommand{\LimitElectronG}{1.51}

\newcommand{\LimitMuonExpectedG}{1.44}
\newcommand{\LimitMuonG}{1.45}

\newcommand{\LimitCombinedGOne}{0.71}
\newcommand{\LimitCombinedGThree}{1.03}
\newcommand{\LimitCombinedGFive}{1.33}
\newcommand{\LimitCombinedExpectedG}{1.63}
\newcommand{\LimitCombinedG}{1.63}

\newcommand{\LimitCombinedSq}{1.60}
\newcommand{\LimitCombinedN}{1.52}
\newcommand{\LimitCombinedPsi}{1.49}
\newcommand{\LimitCombinedChi}{1.64}
\newcommand{\LimitCombinedEta}{1.54}
\newcommand{\LimitCombinedI}{1.56}

\begin{document}

\title{Searches for high mass dilepton resonances in pp collisions at $\sqrt{s}$ = 7 TeV with the ATLAS Experiment}


\author{D. Olivito, on behalf of the ATLAS collaboration}
\affiliation{Department of Physics and Astronomy, University of Pennsylvania, Philadelphia, PA, USA}

\begin{abstract}
The ATLAS detector has been used to search for high mass \epem\ or \mumu\ resonances, such as new heavy neutral gauge bosons. Over 1~\ifb\ of proton-proton collisions at $\sqrt{s}$~=~7~\tev\ recorded by the ATLAS experiment at the Large Hadron Collider are used to search for a high mass state decaying to dilepton pairs.  No excess over Standard Model expectations is observed, and limits are placed on benchmark models predicting spin-1 and spin-2 resonances.
\end{abstract}

\maketitle

\thispagestyle{fancy}


%%%%%%%%%%%%%%%%%%%%%%%%%%%%%%%%%%
\section{Introduction}

High mass resonances are a natural place to look for new physics, and leptonic decays of these resonances would produce striking experimental signatures.  The ATLAS detector has been used to search for high mass resonances decaying to leptons, where lepton in this note refers to an electron or muon.  This analysis uses 1.08~\ifb\ for the \epem\ channel and 1.21~\ifb\ for the \mumu\ channel, recorded in the first half of 2011~(\cite{atlasZprime}).  

%%%%%%%%%%%%%%%%%%%%%%%%%%%%%%%%%%
\section{Theoretical Motivation}

New high mass resonances can arise in a number of models.  In this note, limits are set on spin-1 and spin-2 resonances.  

The benchmark model for spin-1 resonances is the Sequential Standard Model (SSM) \zp\ boson~(\cite{Langacker:2008yv}), which shares the same production as the Standard Model \z, as well as the same leptonic couplings.  Also considered are $E_{6}$ grand unified theory models~(\cite{London:1986dk}), where the $E_{6}$ group is broken down to $SU(5)$ and two $U(1)$ groups.  The neutral gauge fields of the the $U(1)$ groups, $\psi$ and $\chi$, mix to form the \zp\ candidate.  The choice of mixing angle (\te6, defined in Eqn.~\ref{eq:e6}) determines the \zp\ coupling to fermions.
\begin{equation}
\zp (\te6 ) = \zp _\psi \cos{\te6 } + \zp _\chi \sin{\te6 }
\label{eq:e6}
\end{equation}

The benchmark model for spin-2 resonances is the Randall-Sundrum (RS) model~(\cite{RS}), which predicts excited Kaluza-Klein modes of the graviton that appear as spin-2 resonances.  These modes have a narrow intrinsic width when $k/\overline{M}_{Pl}<$ 0.1, 
where $k$ is the spacetime curvature in the extra dimension,  and $\overline{M}_{Pl}  = M_{Pl}/\sqrt{8\pi}$ is the reduced Planck scale.

%%%%%%%%%%%%%%%%%%%%%%%%%%%%%%%%%%
\section{ATLAS Detector}

The ATLAS detector is composed of an inner detector immersed in a solenoidal magnetic field at the innermost radius, electromagnetic and hadronic calorimeters, and a muon spectrometer immersed in a toroid magnetic field furthest from the interaction point~(\cite{atlas:detector}).  The inner detector provides tracking of charged particles for pseudorapidity $|\eta| < 2.5$.  It consists of a silicon pixel detector closest to the beam pipe, a silicon strip detector, and a hybrid transition radiation detector and straw tube tracker at the outermost radius.  The calorimeters span $|\eta| < 4.9$, but for this analysis, the finely-segmented central region covering $|\eta| < 2.47$ is used.  A liquid-argon electromagnetic calorimeter provides precision measurements for electrons in this region.  Outside the calorimeters, air-core toroids provide the magnetic field for the muon spectrometer.  Three sets of precision chambers (Monitored Drift Tubes in the barrel region, Cathode Strip Chambers in the endcaps) provide coverage for $|\eta| < 2.7$.  Trigger chambers provide coverage for $|\eta| < 2.4$, with resistive-plate chambers in the barrel and thin-gap chambers in the endcaps.

%%%%%%%%%%%%%%%%%%%%%%%%%%%%%%%%%%
\section{Event Selection}

The goal of this analysis is to search for resonances in the high invariant mass (\mll) region for electrons and muons.  Thus the event selection requires 2 leptons of like flavor, and selection criteria are applied to the leptons both to minimize the reducible backgrounds and to ensure well-understood leptons at very high transverse momentum (\pt).  Both electrons and muons are triggered with single lepton triggers, with a transverse energy (\et) threshold of 20~\gev\ for electrons and \pt\ threshold of 22~\gev\ for muons.  For the analysis, both electron \et\ and muon \pt\ are required to be greater than 25~\gev, ensuring maximal trigger efficiency.  Event candidates are required to have a reconstructed primary vertex with at least 3 charged particle tracks with $\pt > 0.4$~\gev, to reject cosmic ray events and potential beam backgrounds.

In addition to the above, electron candidates are required to be within $|\eta| < 2.47$ and are required to be outside the calorimeter transition region of $1.37 < |\eta| < 1.52$, which is not as well understood.  Quality cuts are applied to remove electrons reconstructed in areas of the calorimeter with readout issues.  As electrons are reconstructed from an electromagnetic calorimeter cluster matched to an inner detector track, quality requirements are made on both the transverse shower profile in the calorimeter and on the track.  Further requirements are made on the matching between the two.  These correspond to the \emph{Medium} level of ATLAS electron identification~(\cite{egnote}) and provide rejection against background arising from QCD jets.  A hit in the innermost pixel layer is required to reduce the contribution of converted photons.  The leading electron is also required to be isolated in the calorimeter, with $\Sigma \et(\Delta R<0.2)<7\gev$.  The calorimeter cells corresponding to the reconstructed electron are removed from the isolation sum, and corrections are applied for residual leakage of the electron energy as well as other pp interactions in the event of interest (i.e. pileup).  The total acceptance for a \zpee\ with mass 1.5~\tev\ is 65\%, including detector acceptance and efficiencies for trigger, reconstruction, and identification.  The electron energy resolution is effectively flat at high \et, dominated by a constant term of 1.2\% in the barrel and 1.8\% in the endcaps.

Muon candidates consist of a track found in the inner detector and an independent track found in the muon spectrometer.  These tracks are combined in a single fit to determine the momentum of the muon candidate.  To ensure high track quality, hit requirements are made in the inner detector subsystems.  Specific to these high \pt\ muons, the muon candidate is required to have hits in each of three sets of precision chambers in the muon spectrometer (inner, middle, and outer), as the momentum resolution is best understood for these candidates.  Muon candidates are rejected if their tracks have hits in both the barrel and endcap region of the muon spectrometer, to minimize the impact of residual misalignments between subsystems.  These cuts effectively restrict the geometrical acceptance.  Backgrounds are suppressed by additional requirements: to remove contributions from cosmic rays, the muon tracks must have a transverse impact parameter consistent with the primary vertex, $|d_{0}| < 0.2$~mm.  In the longitudinal direction along the beamline, the muon track is required to be within $|z_{0}| < 1$~mm of the primary vertex, and the position of the primary vertex relative to the interaction point must be within $|z(PV)| < 200$~mm.  QCD jet background is suppressed by requiring each muon to be isolated in the inner detector, such that $\Sigma\pt(\Delta R<0.3)/\pt(\mu)<0.05$, where the sum includes tracks with $\pt > 1$~\gev.  Furthermore, since the muon momentum measurement is taken from the curvature of the track, and a muon with misreconstructed charge would have grossly mismeasured momentum, the two muon candidates are required to have opposite charge.  With these requirements, the total acceptance for a \zpmumu\ with mass 1.5~\tev\ is 40\%, again including detector acceptance as well as efficiencies for trigger, reconstruction, and identification.  The lower acceptance relative to the electron channel is largely a result of the track quality requirements imposed in the muon spectrometer, which are necessary for a good understanding of the momentum resolution.  Work is currently ongoing to understand the remaining regions of the muon spectrometer sufficiently well to extend this acceptance.

%%%%%%%%%%%%%%%%%%%%%%%%%%%%%%%%%%
\section{Backgrounds \label{sec:bkg}}

After the selections outlined in the previous section, the primary remaining backgrounds are from Standard Model sources of dileptons.  The main and irreducible background for both channels is Standard Model Drell-Yan production (\zgstar), which yields the same final state as the considered signal models.  Smaller contributions come from electroweak diboson production ($WW$, $WZ$, and $ZZ$), as well as \ttbar\ pairs decaying leptonically.  In the electron channel, QCD jets can fake a prompt electron, through either a photon conversion, semi-leptonic decay of a $b$ or $c$ quark, or a hadron misreconstructed as an electron.  In the muon channel, QCD jets primarily give rise through the semi-leptonic decays of a $b$ and $c$ quarks.  Muons from light mesons decaying in flight are negligible at high \pt.  For both electrons and muons, jets enter the selection through either QCD dijet production or production in association with a \w\ boson (\wpjet).  Cosmic ray background for the muon channel was studied in data and found to be negligible.  

All backgrounds except QCD dijets are taken from simulation.  
The \zp, \gstar\ signal and $Z/\gamma^*$ processes are generated with
\pythia\ 6.421~(\cite{Sjostrand:2006za}) using MRST2007 LO*~(\cite{mrst}) parton distribution functions (PDFs).
Interference between the $Z/\gamma^*$ processes and the heavy resonances is small and therefore neglected.
The diboson processes are generated with \herwig~6.510~(\cite{herwig}) using MRST2007 LO* PDFs.
The \wpjet\ background is generated with \alpgen~(\cite{Mangano:2002ea}) and
the \ttbar\ background with \mcatnlo~3.41~(\cite{mcatnlo}).
For both, \jimmy~4.31~(\cite{jimmy}) is used to describe multiple parton interactions and
\herwig\ to describe the remaining underlying event and parton showers, using CTEQ~(\cite{Pumplin:2002vw}) PDFs.
Final-state photon radiation is handled with \photos~(\cite{fsr_ref}). 
The generated samples are processed through a full simulation of the ATLAS detector~(\cite{atlas:sim}) based on GEANT4~(\cite{geant}).

The $Z/\gamma^*$ cross section is calculated at next-to-next-to-leading order (NNLO)
using PHOZPR~(\cite{Hamberg:1990np}) with MSTW2008 parton distribution functions~(\cite{mstw}). 
The ratio of this cross section to the leading-order cross section is used to determine a mass dependent QCD K-factor which is 
applied to the results of the leading-order simulations. 
The same QCD K-factor is applied to the \zp\ signal. No 
QCD K-factor is available for  \gstar\ production at $\sqrt{s} = 7$~TeV.  
 Higher-order weak corrections (beyond the photon radiation included in the simulation) 
are calculated using \horace~(\cite{horace,CarloniCalame:2007cd}), yielding a weak K-factor due to virtual heavy gauge boson loops.
The weak K-factor is only applied to the Drell-Yan background.
The diboson cross section is known to next-to-leading order (NLO) with an uncertainty of 5\%.
The \wpjet\ cross section is 
rescaled to the inclusive NNLO calculation,
resulting in 30\% uncertainty when at least one parton with $E_{T} > 20$~\gev\ accompanies the \w\ boson.
The \ttbar\ cross section is  predicted at approximate-NNLO, with 10\% uncertainty~(\cite{Moch:2008qy,Langenfeld:2009tc}).

Figure~\ref{fig:pt} shows the electron \et\ and muon \pt\ distributions for selected leptons, with the sum of expected Standard Model backgrounds.  The open histograms show the expected contributions from three \zpssm\ mass hypotheses.  Good agreement is observed between data and expectation.  The excess of data at high \pt\ for muons is understood to be a combination of muon momentum resolution and poor \pythia\ modeling of high \pt\ \zpjet\ events.  The invariant mass \mll\ is modeled well by \pythia\ after the mass dependent QCD K-factor correction, however, as will be seen in Fig.~\ref{fig:mll}.

\begin{figure}[ht]
\centering
\includegraphics[width=.45\textwidth]{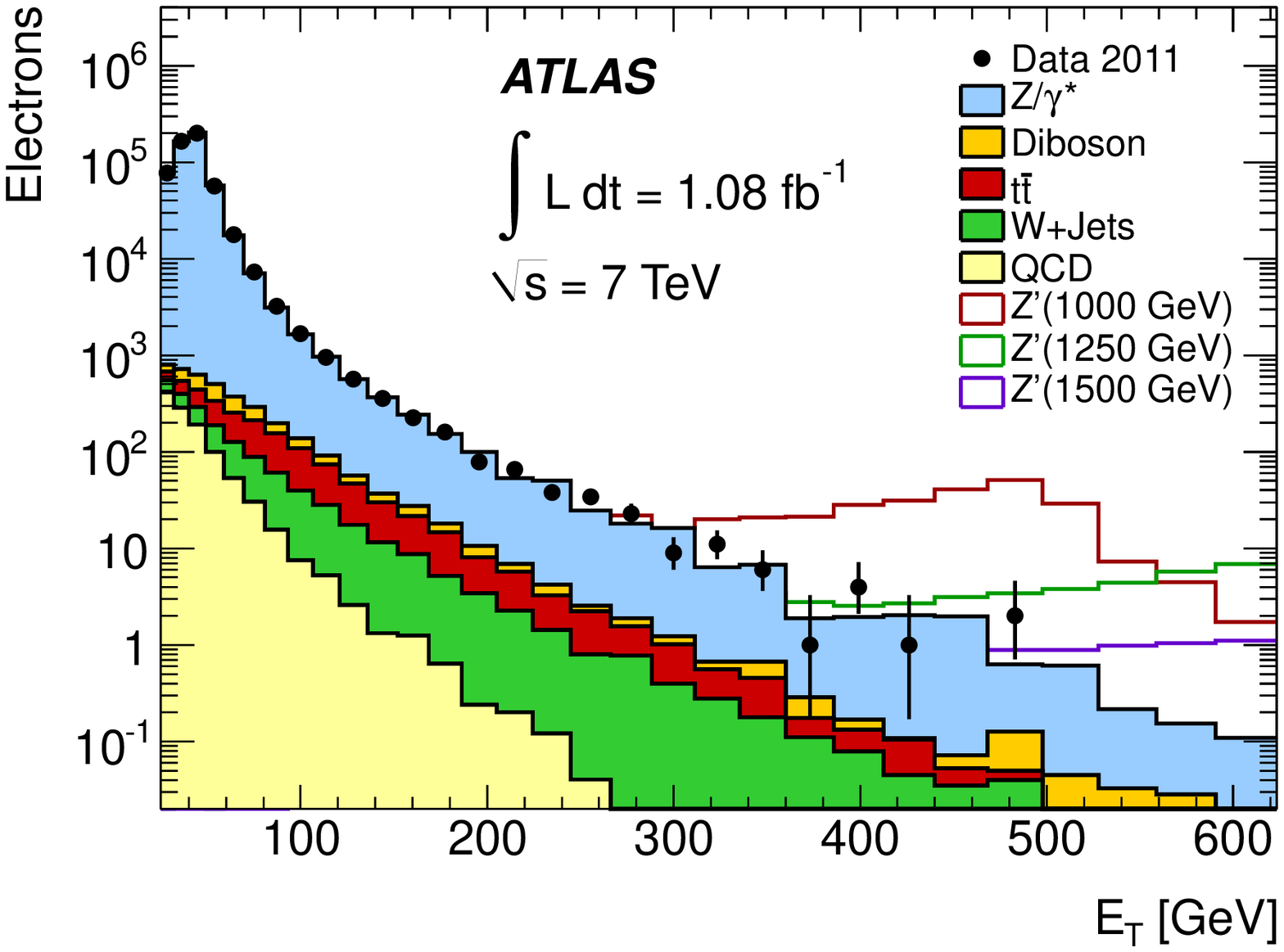}
\includegraphics[width=.45\textwidth]{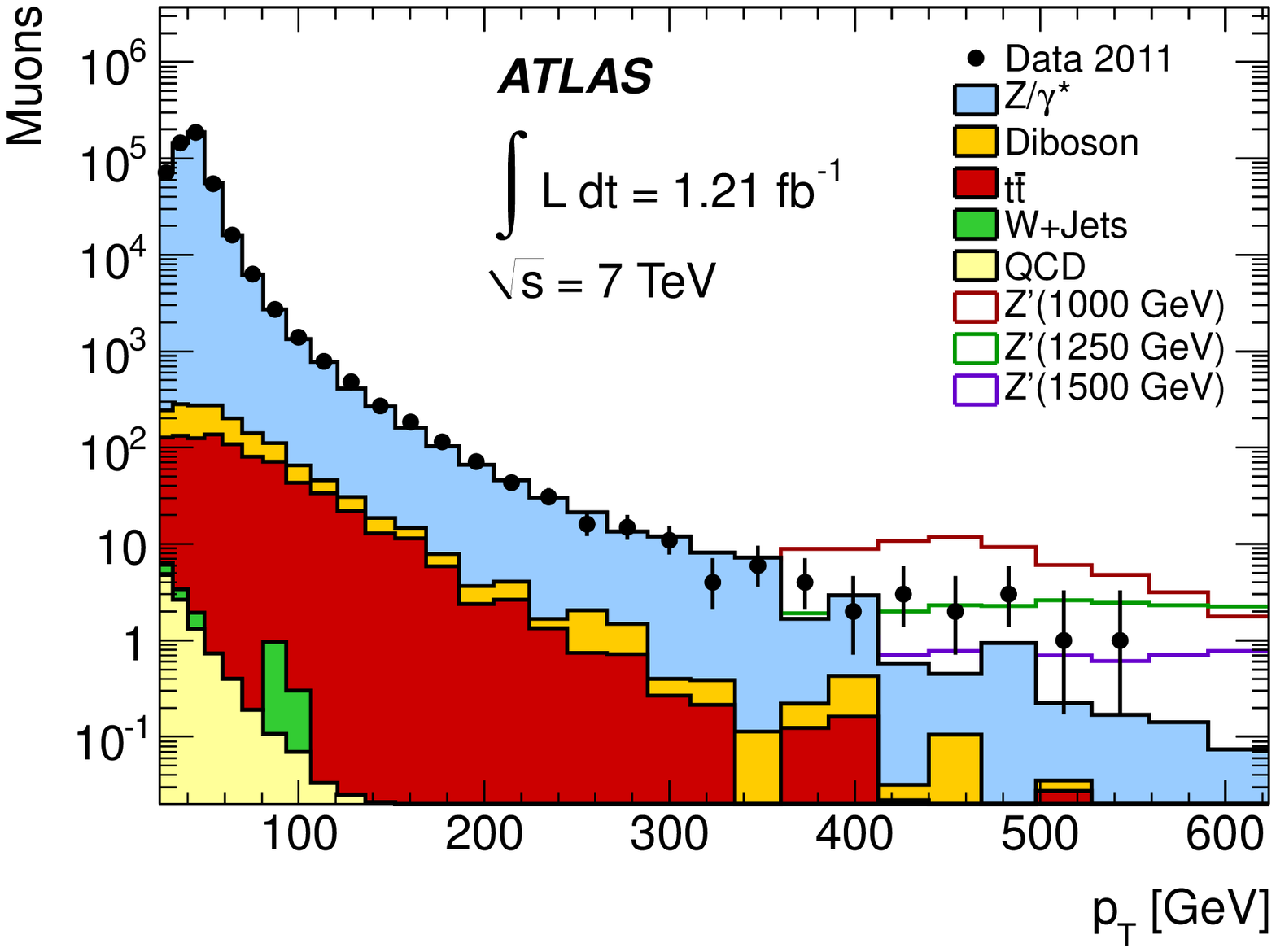}
\caption{Electron \et\ (left) and muon \pt\ (right) for selected leptons, compared with the expected Standard Model background.  The open histograms show the expected contributions from three \zpssm\ mass hypotheses.} \label{fig:pt}
\end{figure}

%%%%%%%%%%%%%%%%%%%%%%%%%%%%%%%%%%
\subsection{QCD Dijets}

Due to poor modeling and low statistics in Monte Carlo, QCD dijet backgrounds are evaluated from data for both channels.  In the electron channel, where dijet background is more significant, three independent data-driven methods are used.  The baseline method is called ``inverted identification.''  In it, a QCD dijet sample is selected in data by requiring two electron candidates which pass looser electron identification criteria but fail the \emph{Medium} criteria used for the analysis selection.  As the statistics in this sample run out before reaching the high mass signal region, the \mee\ distribution from this sample is fit with an empirical function to allow for extrapolation.  To determine the normalization, a two-component fit in \mee\ is performed in the range $70 < \mee < 200$~\gev, where one component is the QCD dijet function and the other is the sum of all backgrounds from simulation (each weighted by cross section).

Two other methods provide cross checks and systematic uncertainties for this estimation.  The first of these, the ``fake rate'' method, derives a fake rate from a dijet sample triggered by QCD jets.  To predict the background in the signal region, it applies this fake rate to a dielectron selection with one electron candidate passing the analysis cuts and the second passing a dijet-enriched selection.  The last method, ``isolation fits,'' uses a binned likelihood fit in calorimeter isolation to distinguish electrons from background.  Since the isolation energy of jets increases at high \et, this method becomes increasingly effective, as long as statistics are sufficient.  Templates are taken from single electron data: the signal template comes from electrons from \w\ boson decays, while the background template comes from a jet-enriched selection where identification cuts are reversed.  As seen in Fig.~\ref{fig:e_iso}, the calorimeter isolation is also fairly well modeled by simulation.  The leading and subleading electron candidates are fit separately.  These results are then combined using a system of equations, to avoid double counting of background events.  

\begin{figure}[ht]
\centering
\includegraphics[width=.45\textwidth]{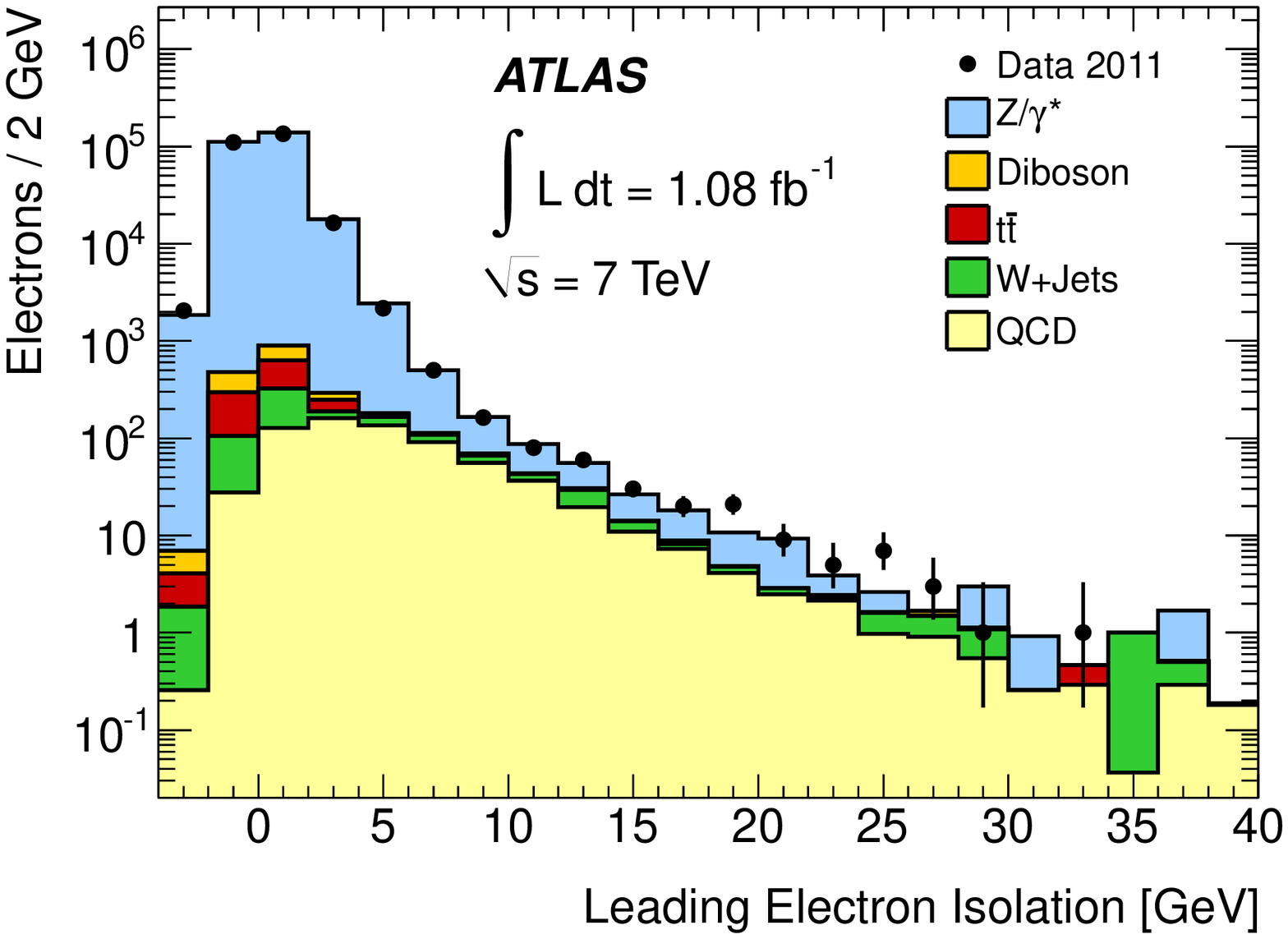}
\includegraphics[width=.45\textwidth]{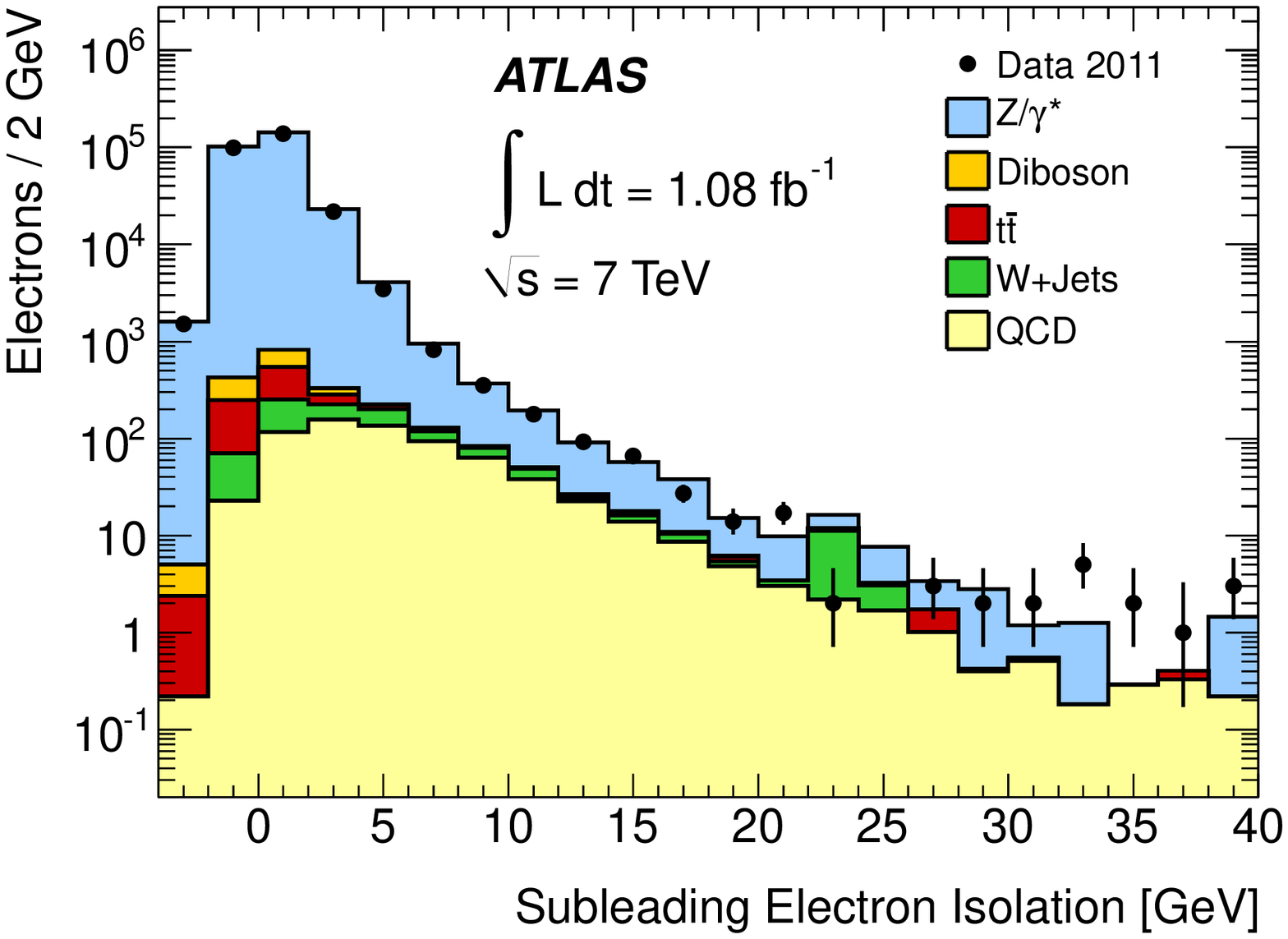}
\caption{Calorimeter isolation for leading (left) and subleading (right) electrons.} \label{fig:e_iso}
\end{figure}

In the muon channel, the QCD background is much smaller but still evaluated from data.  A ``reverse isolation'' method is employed: a QCD sample is selected in data by requiring two muons with $0.1<\Sigma\pt(\Delta R<0.3)/\pt(\mu)<1.0$.  The normalization for this sample is obtained from the ratio of isolated to non-isolated events in QCD \ccbar\ and \bbbar\ Monte Carlo.  As shown in Fig.~\ref{fig:mu_iso}, this region is dominated by QCD background and well described by simulation.

\begin{figure}[ht]
\centering
\includegraphics[width=.45\textwidth]{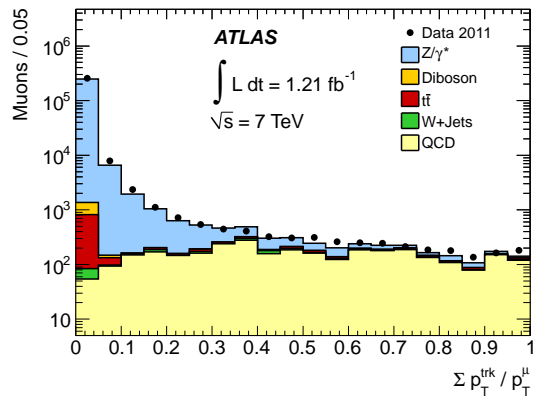}
\caption{Normalized track isolation for muons.} \label{fig:mu_iso}
\end{figure}

%%%%%%%%%%%%%%%%%%%%%%%%%%%%%%%%%%
\section{Results}

Figure~\ref{fig:mll} shows the resulting \mll\ distributions for electrons and muons, compared with the sum of Standard Model expectations.  The open histograms show the expected contributions from three \zpssm\ mass hypotheses.  No clear excess is observed over the background expectations.  For these plots and the limit setting procedure, the sum of simulated backgrounds is normalized to data in the \z\ peak region of $70 < \mll < 110$~\gev, for reasons explained below.  Table~\ref{tab:backgroundTableCombo} shows the background expectations and observed data in bins of \mll.  In the first bin, data and expectation agree by construction.  The QCD dijet background falls off more steeply than the others and is negligible in the highest \mll\ bin for both channels.

\begin{figure}[ht]
\centering
\includegraphics[width=.45\textwidth]{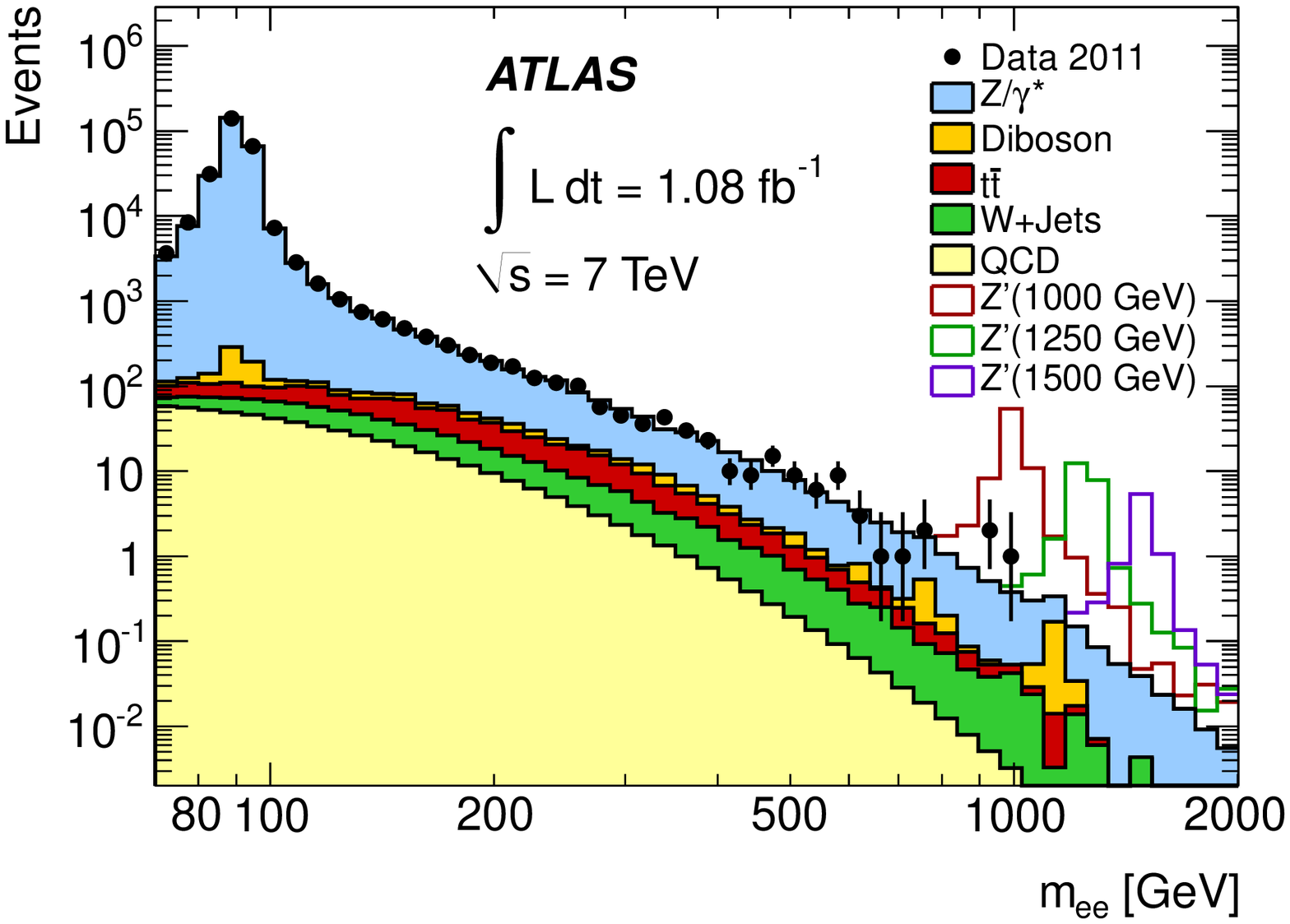}
\includegraphics[width=.45\textwidth]{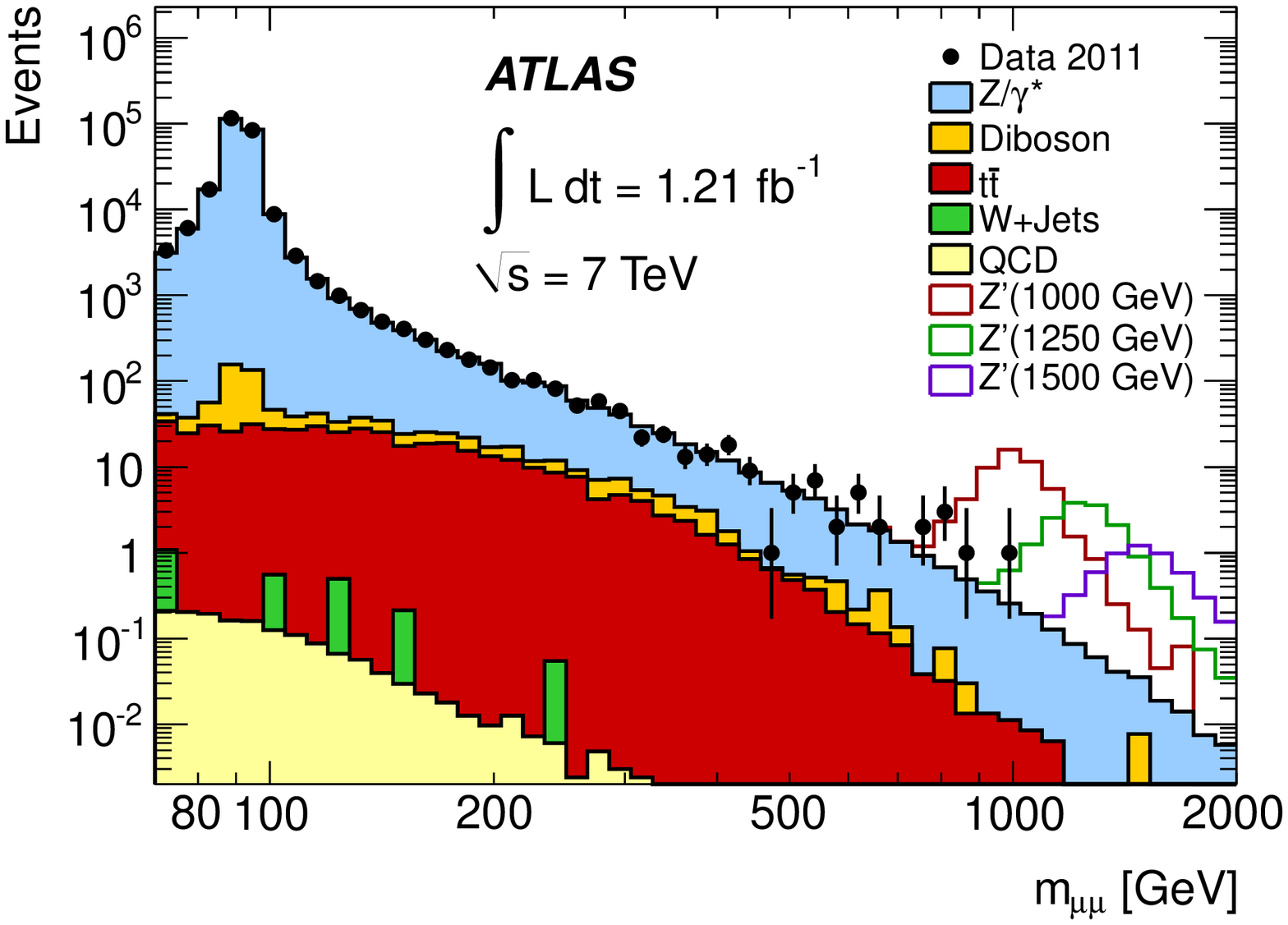}
\caption{Invariant mass (\mll) for selected pairs of electrons (left) and muons (right), compared with the expected Standard Model background.  The open histograms show the expected contributions from three \zpssm\ mass hypotheses.} \label{fig:mll}
\end{figure}

\begin{table}[!htb]
\caption{Expected and observed number of events in the dielectron (top) and dimuon (bottom) channels. 
The first bin is used to normalize the total background to the data. 
The errors quoted include both statistical and systematic uncertainties, except the error on the total background in the normalization region 
which is given by the square root of the number of observed events. 
The systematic uncertainties are correlated across bins and are discussed in the text.  
}
\label{tab:backgroundTableCombo}
\vspace{5mm}
\centering
\small\addtolength{\tabcolsep}{+2pt}
%\resizebox{\columnwidth}{!}{
\begin{tabular}{cccccc}
\hline
\mee [\gev]& 70-110& 110-200  & 200-400 & 400-800 & 800-3000     \\
\hline
DY & 258482 $\pm$ 410 & 5449 $\pm$ 180 & 613 $\pm$ 26 & 53.8 $\pm$ 3.1 & 2.8 $\pm$ 0.1 \\
\ttbar & 218 $\pm$ 36 & 253 $\pm$ 10 & 82 $\pm$ 3 & 5.4 $\pm$ 0.3 & 0.1 $\pm$ 0.0 \\
Diboson & 368 $\pm$ 19 & 85 $\pm$ 5 & 29 $\pm$ 2 & 3.1 $\pm$ 0.5 & 0.3 $\pm$ 0.1 \\
W+jets & 150 $\pm$ 100 & 150 $\pm$ 26 & 43 $\pm$ 10 & 4.6 $\pm$ 1.8 & 0.2 $\pm$ 0.4 \\
QCD & 332 $\pm$ 59 & 191 $\pm$ 75 & 36 $\pm$ 29 & 1.8 $\pm$ 1.4 & $< 0.05$ \\
\hline
 Total & 259550 $\pm$ 510 & 6128 $\pm$ 200 & 803 $\pm$ 40 & 68.8 $\pm$ 3.9 & 3.4 $\pm$ 0.4 \\
\hline
 Data & 259550 & 6117 & 808 & 65 & 3 \\
\hline
%\medskip 
\\

\\ 

\hline
$m_{\mu^{+}\mu^{-}}$ [\gev]  &  70-110  &  110-200  &  200-400  &  400-800  &  800-3000  \\
\hline
DY       &  236319 $\pm$ 320  &  5171 $\pm$ 150  &  483 $\pm$ 22  &  40.3 $\pm$ 2.5  &  2.0 $\pm$ 0.3  \\ 
\ttbar    &  193 $\pm$ 21  &  193 $\pm$ 20  &  63 $\pm$ 6  &  4.2 $\pm$ 0.4  &  0.1 $\pm$ 0.0  \\ 
Diboson  &  307 $\pm$ 16  &  69 $\pm$ 5  &  25 $\pm$ 2  &  1.7 $\pm$ 0.5  &  $< 0.05$  \\ 
\w+jets   &  1 $\pm$ 1  &  1 $\pm$ 1  &  $< 0.5$  &  $< 0.05$  &  $< 0.05$  \\ 
QCD      &  1 $\pm$ 1  &  $< 0.5$  &  $< 0.5$  &  $< 0.05$  &  $< 0.05$  \\ 
\hline
Total    &  236821 $\pm$ 487  &  5434 $\pm$ 150  &  571 $\pm$ 23  &  46.1 $\pm$ 2.6  &  2.1 $\pm$ 0.3  \\ 
\hline
Data     &  236821  &  5406  &  557  &  51  &  5  \\ 
\hline
% \hline
\end{tabular}
%}
\end{table}

%%%%%%%%%%%%%%%%%%%%%%%%%%%%%%%%%%
\section{Uncertainties}

Since the expected backgrounds are normalized to the observed data in the \z\ peak region, luminosity and other uncertainties which are independent of \mll\ (e.g. efficiencies) will cancel out between the \z\ and \zp.  The remaining uncertainties are primarily theoretical.  The 5\% uncertainty on the \z\ cross section is taken as a normalization uncertainty.  At higher \mll, the theoretical uncertainties on the \zgstar\ cross section grow.  The overall uncertainty due to PDF and $\alpha_S$ variations is estimated to be 10\% at 1.5~\tev\ using the MSTW 2008 
eigenvector PDF sets and other PDF sets corresponding to variations of $\alpha_S$.
The difference with respect to CTEQ is included as an additional 3\% uncertainty.
The uncertainty on the QCD K-factor is 3\%, evaluated from variations of the renormalization and factorization
scales by factors of two around the nominal values. A systematic uncertainty of 4.5\% is attributed to EW corrections~(\cite{Aad:2011xp}).

Experimental effects on efficiency, scale, and resolution were studied in data by using leptons from \z\ bosons to compare with simulation, then simulation was used to extrapolate these effects to the high \mll\ region.  For electrons, energy scale and resolution were studied by comparing the \z\ line shape in data and simulation, and the resulting uncertainties are negligible for this analysis.  Electron efficiencies for trigger, reconstruction, and identification were studied using the tag and probe method, resulting in an uncertainty that changes negligibly with \mee.  An additional 1.5\% identification uncertainty is applied to account for potential inefficiencies at high mass due to detector and radiative effects on the calorimeter isolation.

For muons, the resolution is studied by comparing the \z\ line shape from data with simulation, and by comparing muon spectrometer and inner detector tracks.  Simulation was adjusted to the data observations from this low mass region, and the derived uncertainty has an effect of less than 1.5\% on the event yield.  Trigger, reconstruction, and identification efficiencies were studied using the tag and probe technique.  The combined uncertainty is estimated to be 4.5\% at a mass of 1.5~\tev, dominated by a conservative estimate of energy loss from muon bremsstrahlung in the calorimeters.  

The dominant systematic uncertainties are collected in Table~\ref{tab:systematicSummary}, with uncertainties less than 3\% neglected.  No theoretical uncertainties are applied to the signal models considered.

\begin{table}[!t]
\caption{Summary of the dominant systematic uncertainties on the expected signal and background yields at $\mll=1.5$~\tev\ for the \zp\ (\gstar) analysis. NA means not applicable. 
}
\label{tab:systematicSummary}
\centering
\addtolength{\tabcolsep}{+1pt}

%\resizebox{\columnwidth}{!}{
%\tiny
\begin{tabular}{lcccc}
% \hline
\hline
Source        & \multicolumn{2}{c}{dielectrons} & \multicolumn{2}{c}{dimuons} \\
	      & signal & background	   & signal & background \\
\hline
Normalization & 5\%		    & NA		      & 5\%	    & NA \\
PDFs/$\alpha_S$ & NA		    & 10\%		      & NA	    & 10\% \\
QCD K-factor  & NA	    & 3\%		      & NA	    & 3\% \\
Weak K-factor & NA		    & 4.5\%		      & NA	    & 4.5\% \\
Trigger/Reconstruction & negligible	    & negligible  & 4.5\%   & 4.5\% \\
%Resolution    & negligible		    & negligible		       & 1.5\% & negligible \\
%QCD Background & NA		    & 1.5\%		      & NA     & negligible \\
\hline
 Total        & 5\%		& 11\%  		 & 7\%  	& 12\%\\
\hline
% \hline
\end{tabular}
%}
\end{table}

%%%%%%%%%%%%%%%%%%%%%%%%%%%%%%%%%%
\section{Statistical Method}

A template fit method is used to quantify the agreement of the observed data with signal and background expectations.  Signal mass hypotheses from 0.13-2.0~\tev\ were considered.  For each signal hypothesis, a likelihood function is defined as the product of the Poisson probabilities for all \mll\ bins in the search region, computing the probability to obtain the observed data given the background plus signal.  To quantify the agreement with a background-only hypothesis, a $p$-value is computed, representing the probability of seeing an excess at least as significant as the largest observed in data, in the absence of signal.  A scan is performed over \zp\ cross section and mass, with no significant excesses found.  The electron and muon channels yield $p$-values of 54\% and 24\%, respectively.  

%%%%%%%%%%%%%%%%%%%%%%%%%%%%%%%%%%
\section{Limits}

As no signal is observed, 95\% confidence level (C.L.) limits are set on the number of \zp\ (\gstar) events using a Bayesian approach~(\cite{bayesianMethod}) with a flat prior for the signal cross section.  Systematic uncertainties are incorporated as nuissance parameters and marginalized, with a total effect of $\sim$1\% on the limits set.  Limits on the number of signal events are turned in a limit on the cross section times brancing ratio (\xbr) by scaling by $\xbr(Z \rightarrow ll)$ over the observed number of \z\ events. The expected limits are evaluated with pseudo-experiments containing only background events, calculating the 95\% C.L. upper limit for each pseudo-experiment.  The median of the distribution of limits is taken as the expected limit, and the distribution is used to determine the 68\% and 95\% contours corresponding to the expected limit $\pm1,2\sigma$.  Figure~\ref{fig:limits} shows the expected and observed limits for spin-1 (left) and spin-2 resonances (right), combining the electron and muon channels.  The predicted \xbr\ for the benchmark spin-1 model, the \zpssm, is shown with its theoretical uncertainty (as the width of the curve), along with the \esix-motivated \zp\ models with extremal \xbr.  For spin-2, the RS graviton \xbr\ is displayed for couplings $0.01 < k/\overline{M}_{Pl} < 0.1$, with the theoretical uncertainty on $k/\overline{M}_{Pl}$=0.1.  The observed and expected limits for the benchmark models are given in Table~\ref{tab:limits}, with observed limits for the additional models in Table~\ref{tab:combinedLimits}.

\begin{figure}[ht]
\centering
\includegraphics[width=.45\textwidth]{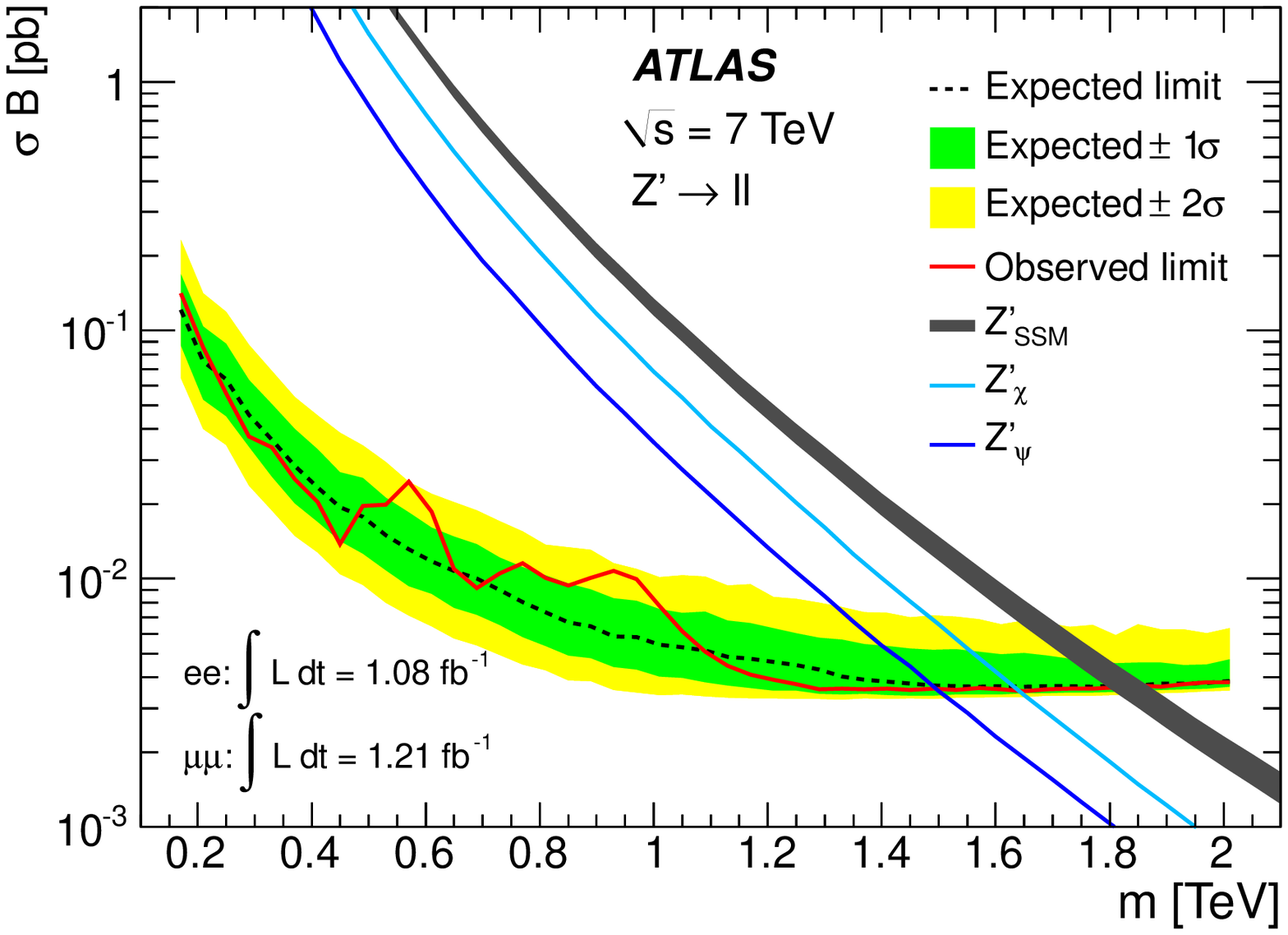}
\includegraphics[width=.45\textwidth]{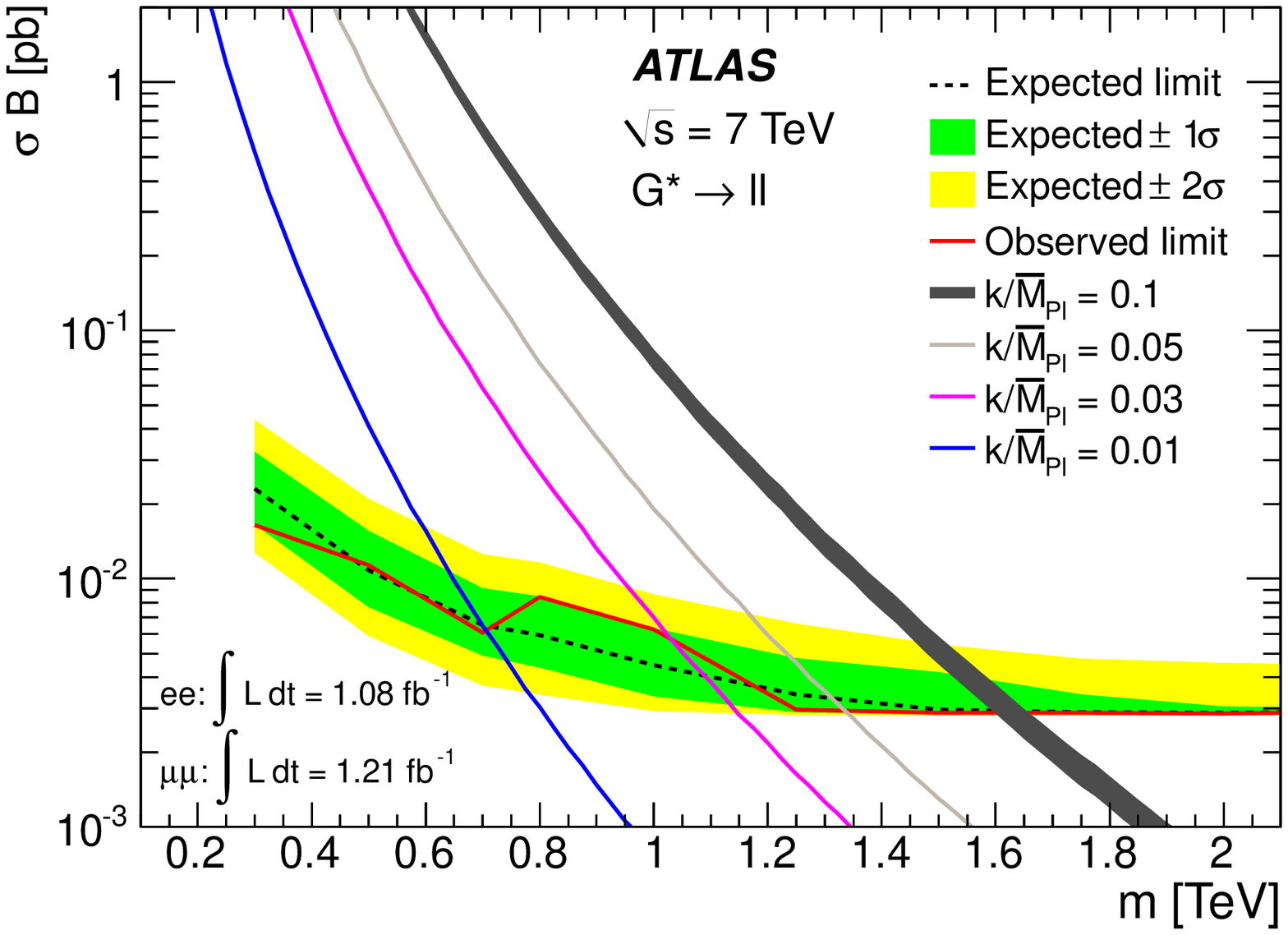}
\caption{Combined limits on $\sigma B$ for spin-1 (left) and spin-2 (right) resonances.  The curves show theoretical cross-sections for various models.} \label{fig:limits}
\end{figure}

\begin{table}[!htb]
\caption{Observed (Expected) 95\% C.L. mass lower limits in \tev\ on \zpssm\ resonance and $G^*$ graviton (with $k/\overline{M}_{Pl}$=0.1).
}
\label{tab:limits}
\begin{center}
%\resizebox{\columnwidth}{!}{
\small
\begin{tabular}{l c c c}
%                         &  Observed mass limit [TeV]            \\
\hline
Model~~~~~          &      $ \ee$~~~      &       $\mumu$~~~  &   $ \ll$~~   \\
\hline			 	 	 	 	  	 
$\zpssm  $                  &   \LimitElectron~(\LimitElectronExpected)  &    \LimitMuon~(\LimitMuonExpected) &   \LimitCombined~(\LimitCombinedExpected) \\
% \hline
$G^* $                     &   \LimitElectronG~(\LimitElectronExpectedG)     &    \LimitMuonG~(\LimitMuonExpectedG) 	    	&	\LimitCombinedG~(\LimitCombinedExpectedG)      \\
\hline
\end{tabular}
\end{center}
\end{table}

\begin{table}[!hbt]
\caption{95\% C.L. lower limits on the masses of \esix-motivated \zp\ bosons and RS gravitons \gstar\ for various values of the coupling $k/\overline{M}_{Pl}$.
Both lepton channels are combined.
}
\label{tab:combinedLimits}
\begin{center}
\small
\begin{tabular}{l|cccccc|cccc}
%\hline
\hline
& \multicolumn{6}{c|}{\esix\ \zp\ Models}   & \multicolumn{4}{c}{RS Graviton} \\
\hline
Model/Coupling            & \zppsi & \zpN  & \zpeta & \zpI  & \zpsq & \zpchi & 0.01 & 0.03  & 0.05 & 0.1 \\
\hline			 	 	 	 	  	 
Mass limit [\tev] & \LimitCombinedPsi  & \LimitCombinedN  & \LimitCombinedEta  & \LimitCombinedI & \LimitCombinedSq & \LimitCombinedChi & \LimitCombinedGOne & \LimitCombinedGThree  & \LimitCombinedGFive  & \LimitCombinedG  \\
\hline
\end{tabular}
\end{center}
\end{table}

%%%%%%%%%%%%%%%%%%%%%%%%%%%%%%%%%%
\section{Conclusions}

Searches were performed for narrow high mass dilepton resonances with the ATLAS detector at the Large Hadron Collider using 1.08-1.21~\ifb\ of data from 2011.  No significant excesses above Standard Model expectations were found, so limits on \xbr\ were set for various models predicting spin-1 and spin-2 resonances.  The 95\% C.L. observed mass limit for the \zpssm\ is \LimitCombined~\tev, while for an RS graviton \gstar\ with coupling $k/\overline{M}_{Pl}$=0.1, the observed limit is \LimitCombinedG~\tev.

\bigskip % extra skip inserted
\bibliography{olivito_zprime_dpf2011}{}

\end{document}